\documentclass[showpacs,pra,twocolumn]{revtex4}
\usepackage{graphicx}
\usepackage{amsmath}
\usepackage{amsthm}
\usepackage{amssymb}
\usepackage{bbm}
\usepackage{color}

\begin{document}

\newcommand{\ketbra}[2]{\ensuremath{\left|#1\right>\left<#2\right|}}
\newcommand{\ket}[1]{\ensuremath{\left|#1\right>}}
\newcommand{\bra}[1]{\ensuremath{\left<#1\right|}}
\newcommand{\braket}[1]{\ensuremath{\left<#1\right>}}
\newcommand{\kb}[1]{\ensuremath{\left|#1\right>\left<#1\right|}}
\newcommand{\bk}[1]{\ketbra{#1}{#1}}
\newcommand{\trace}{\ensuremath{\operatorname{tr}}}
\newcommand{\sign}{\ensuremath{\operatorname{sign}}}
\newcommand{\ie}{\textit{i.e.}}
\newcommand{\eg}{\textit{e.g.}}
\newcommand{\expect}[1]{\left<#1\right>}
\newcommand{\abs}[1]{\left| #1 \right|}
\newcommand{\var}[1]{\ensuremath{\text{Var}( #1 )}}
\newcommand{\mean}[1]{\ensuremath{\left< #1 \right>}}
\newtheorem{proposition}{Proposition}[section]
\newtheorem{problem}{Problem}[section]
\newtheorem{definition}{Definition}[section]
\newtheorem{theorem}{Theorem}[section]
\newcommand{\tr}{\trace}
\newcommand{\WW}{W}
\newcommand{\be}{\begin{equation}}
\newcommand{\ee}{\end{equation}}
\newcommand{\eea}{\end{eqnarray}}
\newcommand{\bea}{\begin{eqnarray}}
\newcommand{\define}{\mathrel{\mathop :}=}
\newcommand{\defineb}{=\mathrel{\mathop :}}
\renewcommand{\vr}{\ensuremath{\rho } }


\title{Iterations of nonlinear entanglement witnesses}
\author{Tobias \surname{Moroder}$^{1,2}$, Otfried
  \surname{G\"uhne}$^{3,4}$, Norbert \surname{L\"utkenhaus}$^{1,2}$} 
\affiliation{$^1$ Quantum Information Theory Group, Institute of
  Theoretical Physics I, and  Max-Planck Research Group for Optics,
  Photonics and Information, University Erlangen-Nuremberg, Erlangen,
  Germany \\ 
  $^2$ Institute for Quantum Computing, University of Waterloo,
  Waterloo, Canada \\ 
  $^3$ Institut f\"ur Quantenoptik und Quanteninformation,
  \"Osterreichische Akademie der Wissenschaften, Innsbruck, Austria\\
  $^4$ Institut f\"ur Theoretische Physik,
  Universit\"at Innsbruck, Technikerstr.~25, A-6020 Innsbruck, Austria}
\date{\today}

\begin{abstract} 
We describe a generic way to improve a given linear entanglement 
witness by a quadratic, nonlinear term. This method  can be iterated, 
leading to a whole sequence of nonlinear witnesses, which become 
stronger in each step of the iteration. We show how to optimize this 
iteration with respect to a given state, and prove that in the limit 
of the iteration the nonlinear witness detects all states that can be 
detected by the positive map corresponding to the original linear witness.
\end{abstract}


\maketitle

\section{Introduction}

When Erwin Schr\"odinger introduced the notion of \emph{entanglement}
for certain bipartite quantum states in the thirties of the last 
century, he might not have imagined that nowadays this peculiar phenomenon 
constitutes the fundamental resource for such fascinating tasks like 
quantum cryptography or quantum teleportation. By definition, an entangled 
state is not separable, which means that it cannot be prepared by local 
operations and classical communication \cite{werner_sep}. Any separable 
state $\rho^{\text{sep}}_{AB}$ can be written as the convex
combination of pure product states, \ie, 
\begin{equation}
  \label{eq: sepstates}
  \rho_{AB}^\text{sep}=\sum_i p_i \ket{\psi_i}_A\bra{\psi_i} \otimes
  \ket{\phi_i}_B\bra{\phi_i}, 
\end{equation}
with a probability distribution $\{ p_i \}_i$ and corresponding pure
states $\ket{\psi_i}_A$ and $\ket{\phi_i}_B$ for the local
subsystems \footnote{Throughout the manuscript only finite dimensional
  systems are considered.}. 

Despite its importance for the field of quantum information theory, the 
properties of entangled states are not fully explored
yet. Even to determine whether a given quantum states is entangled or
not is still an open problem, although considerable progress 
has been achieved along this directions over the last decade, see
Ref.~\cite{horodecki_long}. In fact this so-called \emph{separability
  problem} can already be regarded as a research field on its own and
several different results have provided insight into this problem: the
formulation of operational criteria which are sufficient to detect
either entangled or separable states \cite{peres,
  entanglement_witness, crossnorm, realignment, hofmann, guehne1},
different ways to tackle the problem by numerical means
\cite{eisert_hierarchy, doherty, spedalieri}, or the reformulation of
the separability problem into a different context
\cite{korbicz1,samsonowicz}. Remarkably, in the case of low
dimensionality \cite{entanglement_witness}, or for a particular class
of even infinite dimensional states \cite{simon, werner}, the
separability problem is solved.   

Another approach to the separability problem investigates entanglement
witnesses \cite{entanglement_witness, terhal00a, lewenstein00a}. An
entanglement witness $W$ is an observable which has a non-negative
expectation value on all separable states, and therefore any
negative expectation value signals the presence of entanglement.  
Those kind of operators offer a powerful tool to verify the creation
of an entangled state in an actual experiment, since one only has to
measure the corresponding observable, cf.~Refs.~\cite{guehne_pseudo,
  bourennane04a, lu}.  

\begin{figure}
  \centering
  \includegraphics[scale=0.5,angle=0]{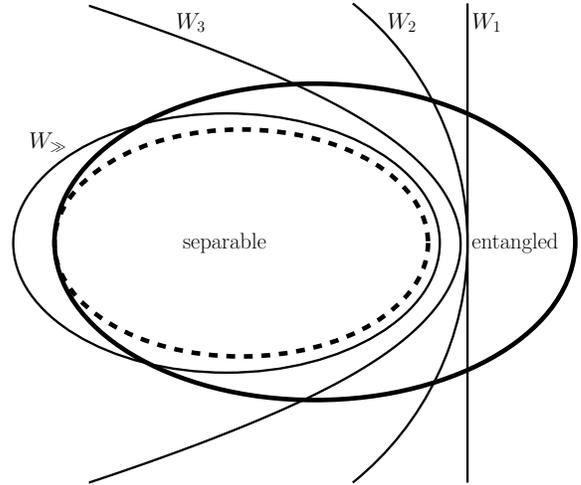}
  \caption{Because the set of separable states (dotted) forms a convex subset
    to the set of all possible density operators (outer, thick-lined), any 
    entangled state can be separated by a corresponding linear
    entanglement witness $\WW_1$. The present manuscript describes an
    iteration method to successively improve entanglement witness,
    corresponding to a certain positive but not completely positive
    map, by quadratic nonlinear terms, and thus generates a sequence
    of nonlinear entanglement witnesses $\WW_2, \WW_3, \dots \WW_{\gg}$ of
    increasing strength.} 
  \label{fig: motivation}
\end{figure}

Witnesses allow the following geometrical interpretation:   
As schematically shown in Fig.~\ref{fig: motivation} the set of all
possible bipartite quantum states forms a convex set, while the
restricted set of only separable states constitutes a convex subset of
it. Therefore it is possible to separate any given entangled state
from the set of separable states by a corresponding
hyperplane. This hyperplane represents the corresponding entanglement
witness, or, more precisely, the states for which the mean value of
the witness equals zero. The witness detects the entangled state,
because its expectation value allows distinction between the states
from either the ``left'' or the ``right'' hand side of the witness.  

From this geometrical picture, it is natural to ask whether it is 
possible to improve a given linear entanglement witness by some 
nonlinear correction in order to approximate the set of separable 
states better. Indeed, as shown in Refs.~\cite{nlwit1,nlwit2} this 
is always possible. In these references a general method has been 
provided to compute a nonlinear improvement to a given linear 
witness. The nonlinear corrections consist of terms which are 
typically squares of certain expectation values, and the whole
expression is still positive on all separable states. The nonlinear 
witnesses may also be viewed as linear observables,  but acting on 
several copies of the input state \cite{nlfuntionals}.

In this paper we provide an extension of this improvement idea which leads 
to a whole sequence of nonlinear witnesses. We first present a 
method to compute a nonlinear improvement for a given witness. 
This method can then be iterated, leading to a sequence of 
nonlinear witnesses. In this iteration, the nonlinear witnesses 
contain higher nonlinearities and become stronger in each step. 
Each iteration step requires a choice of an arbitrary unitary 
transformation. We show how to optimize this choice with respect 
to a given state, in the sense that this state should be detected 
after as few iterations as possible and show that this iteration 
finally detects all the states which are detected by the positive 
map corresponding to the witness.

In detail, our paper is organized as follows. In Sec.~\ref{sec: iteration} 
we present the main idea of the iteration and fix our notation. We consider 
witnesses from the criterion of positivity of the partial transpose
(PPT) \cite{peres} at this point, however, as we will see at the end, the  
methods can directly be used for arbitrary witnesses using the duality 
between witnesses and maps \cite{pillis-67, jamiolkowski, choi-82}. 
In Sec.~\ref{sec: optimized} we discuss the optimization of the iteration 
with respect to a given state and prove its main properties. In 
Sec.~\ref{sec: average} we discuss what happens if the iteration is not
optimized and the unitary transformation is chosen
randomly. Interestingly, it turns out that the resulting nonlinear
witnesses are still strong. In Sec.~\ref{sec: examples} we present
several examples of our iteration. Finally, we discuss extensions of
our method to arbitrary witnesses, and conclude.

\section{Iteration process}
\label{sec: iteration}

The main idea of the iteration process is a particular way of
improving a given entanglement witness by some quadratic, nonlinear
term. The point is that this improvement method has the advantage that
it can be employed several times. This leads to a sequence of
entanglement witnesses of increasing strength. 

In this section we will first introduce the basic idea of the
iteration, and then explain some generic properties of the iteration
process. By giving a simple example sequence, we can show that the
iteration process allows to ``curve'' a given entanglement witness in
such a way that it will detect an arbitrary given state violating the
PPT condition at some point in this sequence.

\subsection{Main idea}
\label{sec: mainidea}

Let us explain the main idea in the simple setting of witnesses coming
from the PPT criterion. For a given density matrix $\rho=\sum_{ij,kl}
\vr_{ij,kl} \ket{ij}\bra{kl}$ the partial transpose with respect to
the second system is given by 
\begin{equation}
  \vr^\Gamma = \sum_{ij,kl} \vr_{il,kj} \ket{ij}\bra{kl}.
\end{equation}
The celebrated PPT criterion states that if \vr is separable, then its
partial transpose is positive semidefinite, \ie, it has no negative
eigenvalues. Hence any quantum state with a negative partial transpose
(NPT) must necessarily be entangled. 

This criterion has two consequences. First, for a separable state we
have $\bra{\phi} \vr^\Gamma \ket{\phi} \geq 0$ for any vector
$\ket{\phi}.$ This is equivalent to $\tr(\vr^\Gamma P) \geq 0$ for any
positive semidefinite operator $P$ or to $\tr(\vr^\Gamma B B^\dagger)
\geq 0$ for any matrix $B$. (See also \footnote{The decomposition
  $P=BB^\dag$ is not unique. For example one can use the
  Cholesky decomposition  \cite{horn85a}, which even restricts the
  operator $B$ to be lower triangular. A different approach consists
  in using the square-root $B=\sqrt{P}=B^\dag$ of $A$, defined by the
  Taylor expansion series, in which each operator $B \geq 0$ is
  positive semidefinite by itself. The main results of the paper are
  formulated for exactly this special class of operators; hence there
  is no restriction of generality.}.) 
Clearly, these conditions are nothing but a reformulation of positive
semidefiniteness of the partially transposed state.  

Second, the PPT criterion allows to construct directly a witness for a
given state violating it. Namely, if $\vr^\Gamma$ has a negative
eigenvalue $\lambda_- < 0$ and $\ket{\psi}$ denotes the corresponding
eigenvector, then   
\begin{equation}
  \label{basicwitness}
  \WW_0 = \ketbra{\psi}{\psi}^\Gamma
\end{equation}
is a witness detecting $\vr$, as $\tr(\vr \WW_0) = \tr(\vr
\ketbra{\psi}{\psi}^\Gamma) = \tr(\vr^\Gamma \ketbra{\psi}{\psi}) =
\lambda_-$, while for all PPT states, thus also separable states, it
holds $\tr(\vr \WW_0)\geq 0$. Note that more generally any observable
of the type $\WW = (B B^\dagger)^\Gamma$ is positive on all PPT
quantum states, and any NPT entangled state can be detected by a
witness of this form. Our aim is to improve this witness by a sequence
of nonlinear witnesses. 

To start, note that $B$ might be a linear combination of some other
operators $\{F_i \}_i$, \ie, $B=\sum_i c_i F_i.$ This implies that for
separable \vr one has 
\begin{equation}
  \trace(\rho^\Gamma BB^\dag) = \sum_{ij} c_i
  \underbrace{\trace(\rho^\Gamma F_i F_j^\dag)}_{\defineb
    M_{ij}(\rho^\Gamma)} c_j^* \geq  0. 
\end{equation}  
As $B$ is not fixed yet, this has to hold for all possible combinations 
of $c_i \in \mathbbm{C}$. Such a condition can be fulfilled if and only if 
the matrix $M(\rho^\Gamma)$ defined above is positive semidefinite
itself. This particular matrix, which we call \emph{expectation value matrix} 
(EVM) in the following, has recently drawn much attention in different
contexts of the quantum information literature, see
Refs.~\cite{korbicz1, vogel, johannes, miranowicz, haeseler, morodertmpsu2}.   


In order to improve the linear entanglement witness, we consider the
EVM for the specific operator set $F_1=U^\dag$ and $F_2=B$, where $U$
constitutes an {\it arbitrary} unitary and the operator $B$
characterizes the linear witness $\WW=(BB^\dagger)^\Gamma.$ Assuming
normalization of the partially transposed state
$\trace(\rho^\Gamma)=1$, the corresponding EVM becomes   
\begin{equation}
  \label{eq: evmit}
  M(\rho^\Gamma)=\left(
    \begin{array}{cc}
      1 & \trace(\rho^\Gamma (BU)^\dag) \\
      \trace(\rho^\Gamma BU) &  \trace(\rho^\Gamma BB^\dag)
    \end{array}
  \right).
\end{equation}
As long as $\vr$ is separable, $\rho^\Gamma \geq 0$ forms a positive
semidefinite operator, and hence the corresponding EVM $M(\rho^\Gamma)
\geq 0$ will be positive semidefinite as well. To finally achieve the
quadratic, nonlinear improvement, one employs the Sylvester criterion
\cite{horn85a}, which states that a given matrix can only be positive
semidefinite if all principle minors are non-negative. This results in
conditions on the diagonal elements and the determinant of the EVM
given by Eq.~\ref{eq: evmit}. Hence the EVM for any separable state
must fulfill 
\begin{eqnarray}
  \label{eq: linwit}
  W_{\text{L}}(\rho)\equiv \trace(\rho^\Gamma BB^\dag) &\geq& 0, \\
  \label{eq: nlwit}
  W_{\text{NL}}(\rho) \equiv\trace(\rho^\Gamma BB^\dag) -
  \left|\trace(\rho^\Gamma B U)\right|^2 &\geq& 0, 
\end{eqnarray}
since otherwise the partially transposed state has a least one negative
eigenvalue and therefore signals the presence of entanglement of the
corresponding quantum state, \footnote{Of course there is another
  principle minor, the entry of the first row and column of the matrix
  given in  Eq.~\ref{eq: evmit}, but since $1 \geq 0$ this will give
  no further conditions.}. 

The first condition given by Eq.~\ref{eq: linwit} represents the
original linear witness $W_\text{L}$. The nonlinear entanglement
witness $W_\text{NL}$ defined by Eq.~\ref{eq: nlwit} is a strict
improvement of the linear witness, as one further subtracts another
non-negative term \footnote{The Sylvester criterion \cite{horn85a} can
  be applied directly to the partially transposed operator $\rho^\Gamma$, which
  results in a finite number of principle minors in the case of finite
  dimensions. Since any principle minor corresponds to a certain
  nonlinear entanglement witness, via the adjoint map,
  cf. Refs. \cite{carteret,vogel, haeseler}, one has a finite number
  of such nonlinear entanglement witnesses that detect all the quantum
  states which are detectable by the partial transpose. This number of
  entanglement witnesses can even be reduced further if one employs another
  connection between the principle minors and non-negativity of the 
  corresponding eigenvalues: such a connection has been for example used
  in Ref. \cite{byrd} and will generate exactly $d-1$ nonlinear entanglement
  witnesses, where $d$ denotes the dimension of the bipartite Hilbert
  space. However in our approach, we likes to have a sequence of
  entanglement witnesses of increasing strength, so any improvement of
  a given entanglement witness must at least detect all the states
  which are ``witnessed'' by the entanglement witness in the step
  before. To our knowledge, this particular ``ordering'' cannot be
  achieved by any  principle minor ordering.}. 

The main point of this construction is that the nonlinear improvement
$W_\text{NL}$ can again be written as an expectation value of a {\it
  linear} witness, and hence the procedure can be iterated. Namely, if
we define 
\begin{equation} 
  \label{eq: bitexp}
  B^\prime \equiv B^\prime(B ,U,\rho^\Gamma) \define B  U  -
  \trace(\rho^\Gamma B U) \mathbbm{1},  
\end{equation}
we have 
\be
W_{\text{NL}}(\rho) = \trace(\rho^\Gamma B^\prime B^{\prime \dag})
\ee
and we can write down a further improvement and so forth. Note that
the defined operator $B^\prime$ depends on the starting operator $B$, the 
chosen unitary $U$ and the considered partially transposed state 
$\rho^\Gamma$.

To give an example, for the case of the witness $\WW_0$ in
Eq.~\ref{basicwitness} we can choose $B=\kb{\psi}$, and an arbitrary
$\ket{\phi}=U^\dag \ket{\psi}$, then the nonlinear entanglement witness becomes
\begin{equation}
  W_\text{NL}(\rho)=\trace(\rho W_0 ) - \big|\trace\big[\rho
  (\ket{\psi}\bra{\phi})^\Gamma\big]\big|^2, 
\end{equation}
which is similar to the quadratic improvement from
Ref.~\cite{nlwit1}. (See also \footnote{However, the particular method
  introduced in Ref.~\cite{nlwit1} enables subtraction of an even
  larger term, but this method can not be iterated.}.)  
 
This iteration poses now a set of interesting questions. First, one
may ask whether one can always find a sequence of improvements of a
witness $\WW,$ which detects a given state $\rho$, which is not yet
detected by the starting witness $\WW.$ As we will see, this is the
case. Second, one may ask how one can find the optimal iteration, such
that the nonlinear term which is subtracted is maximized in each
step. Third, one can investigate what happens if the unitaries are
chosen randomly. We will address all these questions in the following.  

Finally it should be noted that the described iteration process to 
obtain nonlinear witnesses only depends on the knowledge whether 
the partial transposed state is positive semidefinite or not. 
Therefore no entangled quantum state with a positive partial transpose
is detected by any of those nonlinear entanglement witnesses, however
the method can directly be generalized to other positive but not
completely positive maps, cf. Sec. \ref{sec: conclusion}.

\subsection{Definition, properties and example iteration}

The following definition formally introduces the iteration
process. More specifically the iteration contains the sequence of
entanglement witnesses, which are defined via their relation to the
operators $B$ (cf. Sec. \ref{sec: mainidea}), the corresponding
sequence of expectation values and the resulting sequence of
improvements. However, note that the iteration process itself is not
fully specified yet, because so far no explicit recipe to draw the
unitary operators has been fixed. In general, one could distinguish
different kinds of iterations depending on the strategy to choose each
unitary. In the subsequent sections we mainly distinguish three
different iterations: the \emph{optimized iteration} (Sec. \ref{sec:
  optimized}) in which each unitary is chosen such that it maximizes
the improvement for a given quantum state, a \textit{random iteration}
where each unitary is chosen at random and the \emph{averaged
iteration} (Sec. \ref{sec: average}), which considers the averaged
expectation value taken over all possible unitaries.   

\begin{definition}[Iteration process]
\label{def: itproc}
Given the operator $\rho^\Gamma$ with $\trace(\rho^\Gamma)=1$ together
with a starting operator $B_1$, the iteration process contains the
following sequences with $n \geq 1$. Each sequence depends on the set
of unitaries $U_1, \dots U_{n-1}$, which are chosen in the steps
before. 
\begin{itemize}
\item The sequence of \emph{operators} $\{ B_n\}_n$, which are
  recursively defined by  
  \begin{eqnarray}
    \nonumber
    B_n &\equiv& B_n(B_{n-1},U_{n-1},\rho^\Gamma)\\ 
    \label{eq: recB} 
    &\define& B_{n-1} U_{n-1} - \trace(\rho^\Gamma B_{n-1} U_{n-1})\mathbbm{1},   
  \end{eqnarray}
  and the operator $B_1$ is the given starting operator.
\item The sequence of \emph{expectation values} $\{w_n\}_n$, that are
  defined via     
\begin{equation}
  \label{eq: expval}
  w_n \equiv w_n(\rho^\Gamma, B_n) \define \trace(\rho^\Gamma B_n B_n^\dag ). 
\end{equation}
\item The sequence of \emph{improvements} $\{c_{n}\}_n$, which
  characterize the detection improvement in each step, are given by 
\begin{equation}
  c_n \equiv c_n(\rho^\Gamma,B_n, U_n) \define
  \left|\trace(\rho^\Gamma B_n U_n)\right|^2.  
\end{equation}
\end{itemize}
\end{definition}

Note that each operator $B_n\equiv B_n(\rho^\Gamma, U_1, \dots U_{n-1})$,
each expectation value $w_n \equiv w_n(\rho, U_1, \dots U_{n-1})$ and each
improvement $c_n \equiv c_n(\rho^\Gamma, U_1, \dots, U_{n-1})$ depends on the
unitaries $U_1, \dots U_{n-1}$ chosen in the iteration steps before, 
the operator $\rho^\Gamma$ and the given starting operator $B_1$;
however, this explicit dependence is suppressed in most of the
following cases. Even without specifying the exact sequence of
unitary operators, the iteration process has some generic properties:

\begin{proposition}[Generic properties] \label{prop: genprop} For the
  iteration process, introduced by Def.~\ref{def: itproc}, one has the
  following generic properties, which are independent of the starting
  operator $B_1$:   
\begin{itemize}
\item For a given operator $\rho^\Gamma$ that satisfies $\trace(\rho^\Gamma)=1$,
  the sequence of expectation values $\{w_n\}_n$ is monotonically decreasing.
\item If the state is PPT, $\rho^\Gamma \geq 0$, then
  the expectation values $w_n(\rho^\Gamma) \geq 0$ are non-negative,
  for all possible $n\geq 1$. 
\end{itemize}
\end{proposition}
\begin{proof}
Using the recursion formula for $B_n$ and the definition of the $c_n$,
one gets an analogous recursion statement  
\begin{equation}
  \label{eq: recexp}
  w_n=w_{n-1}-c_{n-1}.
\end{equation}
The first statement directly follows from that. The second point
follows from the definition of the expectation value
$w_n=\trace(\rho^\Gamma B_n B_n^\dag)$, and the fact that  $B_n
B_n^\dag \geq 0$  and \mbox{$\rho^\Gamma \geq 0$}.   
\end{proof}

Before considering more specialized situations, we discuss a simple
example iteration. It shows that if one starts with an arbitrary
entanglement witness corresponding to the operator $B_1>0$ it is
always possible to improve the given witness in such a way that it
detects the preselected NPT entangled state.

\begin{proposition}[Example sequence] For a given detectable operator
  $\rho^\Gamma \not \geq 0$ with $\trace(\rho^\Gamma)=1$ and $B_1 > 0$, consider
  the following sequence of unitaries, 
  \begin{equation}
    \label{eq: exampleU}
    U_1=P_+ - P_-, \;\;\; U_n=-P_+ + P_-\;\; \forall n\geq 2,
  \end{equation}
  where $P_-$ denotes the projector onto the subspace of negative
  eigenvalues, and $P_+$ its orthogonal complement. Then there exists an
  $N_0 \in \mathbbm{N}_+$ such that $w_n <0$ for all $n \geq N_0$, hence
  the corresponding sequence of expectation values $\{w_n\}_n$ will
  detect the state; in fact the sequence $\{ w_n \}_n$ diverges to $- \infty$.  
\end{proposition}
\begin{proof}
The recursion formula for the sequence of expectation values $\{ w_n
\}_n$, given by Eq.~\ref{eq: recexp}, allows us to prove the
proposition by showing that the sequence of improvements $\{ c_n \}_n$
does not converge to zero, \ie, $\lim_{n \to \infty} c_n > 0$; in
fact, we even prove divergence of this sequence. Because of the
special set of unitaries, each improvement is given by $c_n =
\left[\trace(\rho^\Gamma B_n U_n)\right]^2$. The first two
improvements can be directly computed, and are given by $c_1=
[\trace(|\rho^\Gamma| B_1)]^2 >0$ and  
\begin{eqnarray}
  \nonumber
  c_2&=&\left[\trace(|\rho^\Gamma| B_1) \| \rho^\Gamma\|_1 -
  \trace(\rho^\Gamma B_1)\right]^2 \\
  \label{eq: c2example}  
  &\geq&  \| \rho^\Gamma \|_1^2
  \left\{\trace[(|\rho^\Gamma|-\rho^\Gamma) B_1] \right\}^2 > 0,
\end{eqnarray} 
where  $\| ... \|_1$ denotes the trace norm. The first inequality is
valid for the case that $\trace(\rho^\Gamma B_1) \geq 0$; if this is
not the case, then $c_2>0$ trivially holds. The strict inequality in
Eq.~\ref{eq: c2example} comes from the fact that
$(|\rho^\Gamma|-\rho^\Gamma)\geq 0$ and the assumption that
$B_1>0$. Note that in the case of a positive semidefinite operator
$\rho^\Gamma=|\rho^\Gamma| \geq 0$, hence the second improvement vanishes. In
the remaining part of the proof, we want to show by induction that      
\begin{equation}
  c_n= c_2 \| \rho^\Gamma\|_1^{2(n-2)}, \;\;\forall n\geq 2. 
\end{equation}
The starting case $n=2$ is trivial, and we only need to care about the
induction step $n \mapsto n+1$. One arrives at  
\begin{eqnarray}
  \nonumber
  c_{n+1}&=& \left[\trace(\rho^\Gamma B_{n}) - \trace(\rho^\Gamma U_{n+1}) 
    \trace(\rho^\Gamma B_n U_n)\right]^2 \\ 
  \nonumber
  &=& \left\{\trace(\rho^\Gamma B_{n-1} U_{n-1}) [1-
      \trace(\rho^\Gamma) ] \right. \\ 
  \nonumber
  && \left. + \| \rho^\Gamma\|_1
    \trace(\rho^\Gamma B_n U_n)\right\}^2\\ 
  &=& c_2 \| \rho^\Gamma\|_1^{2[(n+1)-2]},
\end{eqnarray}
where we used the recursion formula for the operators $B_{n+1}$ and
$U_n U_{n+1}=\mathbbm{1}$ in the first line. In the second step, we
employed the recursion formula for the operator $B_{n}$ and the
identity $\trace(\rho^\Gamma U_{n+1}) = - \| \rho^\Gamma\|_1$, while 
in the last step we used $\trace(\rho^\Gamma)=1$ and the induction
hypothesis in the end. Since $\| \rho^\Gamma\|_1 > 1$, the
corresponding sequence of improvements $\{c_n \}_n$ diverges and the
proposition is proved.   
\end{proof}

This sequence of unitaries can always be ``started'' at each step of
an arbitrary sequence; detection and divergence are ensured as long as
the second expectation value, after one has used this particular
sequence, is strictly greater than zero.

\section{Optimized iteration}
\label{sec: optimized}

Suppose that one starts with a certain entanglement witness and that one
likes to verify the entanglement of a given target state. Then it
is of course desirable to detect the state with the least number of
iterations. One possible way to achieve this goal might be to
try to optimize the improvement for the specific quantum state in each
step. Such a particular iteration method is termed \emph{optimized
  iteration} and is formally introduced in the following
definition. Is is important to note that the sequence of unitaries,
which are chosen during the iteration process, is mainly determined by
the given target state. From a geometrical point of view, this process
corresponds to the task that one wants to curve a given entanglement
witness along a certain direction, and the direction is linked to the
quantum state. The final result of this section is the statement that
this optimized sequence will finally detect the initial target state.

\begin{definition}[Optimized iteration] \label{def: optit} 
  For a given operator $\rho^\Gamma$ with $\trace(\rho^\Gamma)=1$ and
  a certain starting operator $B_1$, the \emph{optimized iteration} is
  defined as the iteration in which each improvement is maximized over
  the chosen unitary. The maximization of  $c_n = |\tr(\vr^\Gamma B_n
  U_n)|^2$ can be obtained as follows: For any valid singular value
  decomposition $\rho^\Gamma B_n=V_n D_n W_n^\dag$ with $D_n \geq 0$,
  one selects the optimal unitary $U_n^{opt}=W_n V_n^\dag,$ resulting in 
  \begin{equation}
    c_n=\left[\trace(\rho^\Gamma B_n U^{\text{opt}}_n)\right]^2 =
    \left[\trace(D_n)\right]^2 = \| \rho^\Gamma B_n \|^2_1. 
  \end{equation} 
  This particular unitary has the additional feature that $\rho^\Gamma
  B_n U_n^{\text{opt}} \geq 0$ is a positive semidefinite operator
  \footnote{The optimality follows from the properties of the singular
    value decomposition, see Theorem 7.4.9 in
    Ref.~\cite{horn85a}. However, there is an ambiguity in this
    definition, since the singular value decomposition is not
    necessarily unique. In such cases, one is free to choose any valid
    decomposition. Note that if one only demands for an optimal
    improvement $\max_{U_n} c_n(U_n)$, the resulting operator
    $\rho^\Gamma B_n U_n^{\text{opt}}$ might not even be hermitian,
    because one could always add an arbitrary random phase to the
    operator $U_n^\text{opt}$.}.  
\end{definition}

The extra requirement that the operator after the optimization
$\rho^\Gamma B_n U_n^{\text{opt}} \geq$ forms a positive semidefinite
operator is of particular importance in order to show detection of an
arbitrary target state, because it allows some predictions of the
structure of the next optimal unitary. The proof itself relies on the
idea that the sequence of improvements does not converge to zero.

\begin{theorem}[Detection for the optimized iteration] For any detectable state
  $\rho^\Gamma \not \geq 0$ with $\trace(\rho^\Gamma)=1$, and any
  strictly positive starting operator  $B_1 > 0$, the optimized
  iteration process will always detect the state.   
\end{theorem}

\begin{proof}
Because of the recursion formula of the sequence of expectation
values, given by Eq. \ref{eq: recexp}, it suffices to prove that
each improvement $c_n$ is bounded from below by a strictly positive
constant. For the optimal iteration method, each improvement
simplifies to $c_n = d_n^2$ with $d_n=\trace(\rho^\Gamma B_n
U_n^{\text{opt}}) \in \mathbb{R}$, hence it suffices to prove this
bound only for the sequence $\{d_n \}_n$. Note that the first
improvement, given by \mbox{$d_1=\trace(\rho^\Gamma B_1
  U_1^{\text{opt}})=\| \rho^\Gamma B_1 \|_1 > 0$}, is strictly
positive. According to the definition of the optimized iteration
process, the optimal unitary in the first step $U_1^{\text{opt}}$ is
determined by a singular value decomposition of the operator
$\rho^\Gamma B_1$ and that one obtains a positive semidefinite
operator $\rho^\Gamma B_1 U_1^{\text{opt}}\geq 0$. In the following,
we want to prove the bound   
\begin{equation}
  \label{eq: boundopt}
  d_n = \trace(\rho^\Gamma B_n U_n^{\text{opt}}) \geq c_1
  \lambda_{\text{min}}> 0,\;\; \forall n \geq 2, 
\end{equation}
where $\lambda_{\text{min}}$ denotes the absolute value of the most
negative eigenvalue of the operator $\rho^\Gamma$. The corresponding
eigenvector is labeled by $\ket{\psi}$.  

The case $n=2$ already covers the main idea. Because of the particular
unitary that we chose in the first step, the next operator
$\rho^\Gamma B_2 = \rho^\Gamma B_1 U_1^{\text{opt}} - d_1 \rho^\Gamma$
is already hermitian. Therefore we can expand the operator in terms of
its spectral decomposition as $\rho^\Gamma B_2 = \sum \eta_i
\ket{v_i}\bra{v_i}$. From this explicit decomposition one can directly
infer the next optimal unitary, which is given by
$U_2^{\text{opt}}=\sum \sign(\eta_i)\ket{v_i}\bra{v_i}$, and
$\sign(\eta_i)$ denotes the sign of the corresponding eigenvalue. As a
result, one obtains a new positive semidefinite operator $\rho^\Gamma
B_2 U_2^{\text{opt}}= \sum | \eta_i | \ket{v_i}\bra{v_i} \geq
0$. These properties allow the following statement,  
\begin{eqnarray}
  \nonumber
  d_2&=& \trace(\rho^\Gamma B_2 U_2^{\text{opt}}) \geq
  \braket{\psi|\rho^\Gamma B_2 U_2^{\text{opt}}| \psi} \\
  \nonumber
   &=& \sum | \eta_i
  | \braket{\psi| v_i}\braket{v_i|\psi} \geq \sum 
  \eta_i \braket{\psi| v_i}\braket{v_i|\psi} \\ 
  \nonumber
  &=& \braket{\psi|\rho^\Gamma B_2| \psi} = \braket{\psi|
      \rho^\Gamma B_1 U_1^{\text{opt}} | \psi} - d_1
  \braket{\psi| \rho^\Gamma| \psi} \\ &\geq& d_1 \lambda_{\text{min}}.
\end{eqnarray}  
In the first step, one uses that the trace over a positive
semidefinite operator is lower bounded by the overlap over only one
possible state. Note that the inequality $\braket{\psi| \rho^\Gamma
  B_2 U_2^\text{opt}|\psi} \geq \braket{\psi| \rho^\Gamma B_2 |\psi}$
only relies on the particular spectral decomposition of the operator
$\rho^\Gamma B_2$ and the explicit choice of the unitary
$U_2^\text{opt}$. Therefore this property holds in any step of the optimized
iteration process. Using this idea multiple times enables us to prove
the general statement,    
\begin{eqnarray}
  \nonumber
  d_n&=& \trace(\rho^\Gamma B_n U_n^{\text{opt}}) \geq \braket{\psi|
    \rho^\Gamma B_{n-1} U_{n-1}^{\text{opt}} | \psi} + d_{n-1}
  \lambda_{\text{min}} \\  
  \nonumber
  &\geq& \braket{\psi| \rho^\Gamma B_{n-2} U_{n-2}^{\text{opt}} |
    \psi} + (d_{n-2}+d_{n-1})  \lambda_{\text{min}} \\ 
  \nonumber
  &\geq& \ldots
  \geq \braket{\psi| \rho^\Gamma B_1 U_1^{\text{opt}} |
      \psi} + \lambda_{\text{min}} (d_1 + \ldots + d_{n-1})
  \\ &\geq& d_1 \lambda_{\text{min}}. 
\end{eqnarray}
This finally proves the bound of Eq.~\ref{eq: boundopt}, and
therefore shows that the general sequence of improvements $\{ c_n \}$ does not
converge to zero. As a consequence the corresponding sequence of expectation
values $\{ w_n \}$ will diverge to $- \infty$, so that there exists a
particular $N_0 \in \mathbbm{N}_+$, such that $w_n < 0$ for all $n\geq
N_0$. Therefore the state $\rho^\Gamma \not \geq 0$ will be detected
at some point in the sequence.   
\end{proof}

This theorem can be extended to the case in which the operator $B_1
\geq 0$ is positive semidefinite and $c_1=\| \rho^\Gamma B_1 \|_1 >0$,
since the method to lower bound each improvement does not rely on
strict positivity of the starting operator. However in the specific
case of $c_1=0$, the partially transposed state $\rho^\Gamma$ and the
starting operator $B_1$ are acting on completely orthogonal subspaces,
and hence the sequence can never detect the state.

\section{Averaged iteration}
\label{sec: average}

Obviously, one drawback of the optimized iteration process is that the
target state has to be known in advance; however in certain cases such
prior knowledge might be unavailable. Therefore, it is interesting
what happens if one chooses the unitaries in the iteration process in
a different, state independent way. A first simple method would be to
choose each unitary at random. At first sight, such a \textit{random
  iteration} seems to produce only small improvements, however if one
repeats the iteration many times one still can hope to detect many
states. We will discuss this random iteration process with an examples
in the next section.  

In the present section we study the \textit{averaged iteration}, in
which, instead of using only a single sequence of unitaries, one takes
the average over \emph{all} possible unitaries in the iteration
process. As before, a negative expectation value for these mean values
signals the presence of entanglement. However, one should mention that the
resulting nonlinear entanglement witnesses can not be written as a
single nonlinear witness in the original form of the iteration. 
  
The final theorem of this section states that there is a sequence of
nonlinear entanglement witnesses, which in the limit of infinite many
improvements, detects all NPT entangled states at once. Although this
results seems surprising at first, such a method is already known
\cite{ekert} and relies on the seminal spectrum estimation method introduced in
Ref.~\cite{keyl}. Nevertheless, a similar statement can be derived
using the idea of the averaged iteration process.

\begin{definition}[Averaged iteration] For a given operator
  $\rho^\Gamma$ that acts on an $d$-dimensional, composite Hilbert
  space, with $\trace(\rho^\Gamma)=1$, and for a certain starting
  operator $B_1$, the \emph{averaged iteration} defines the special
  sequence of expectation values $\{ \overline w_n \}_n$, in which one
  takes the average over all possible unitaries $U_1, \ldots,
  U_{n-1}$, that one can choose up to the $n$-th step. More precisely,
  one defines    
  \begin{equation}
    \label{eq: expave}
    \overline w_n \define \iint dU_1 \ldots d U_{n-1} w_n(\rho^\Gamma,
    B_1, U_1, \ldots, U_{n-1}), 
  \end{equation} 
  with $w_n(\rho^\Gamma, B_1, U_1, \ldots, U_{n-1})$ being the
  expectation value defined by Eq.~\ref{eq: expval}, and where $d U_i$,
  for $i=1,\ldots n$, represents the Haar measure of the unitary group
  $U(d)$ with normalization $\int d U_i =1$. 
\end{definition}

The intuition behind the detection statement of the averaged iteration
process is very simple: One knows already certain sequences of
unitaries (\textit{e.g.} example or optimized iteration) that will
detect a given quantum state and even diverge to $-\infty$. In
contrast for any other sequence of unitaries, in particular those
which never detect the state, the corresponding expectation values are
monotonically decreasing and bounded from above. Hence if one performs
the average over all possible sequences of unitaries, the resulting
sequence of mean values should diverge as well.  

Detection for the averaged iteration process is shown in a two step
procedure: In a first step one obtains a recursion formula for the
sequence of averaged expectation values. This particular formula
simplifies the final proof of detection.

\begin{proposition}[Recursion formula for the averaged iteration]
  \label{prop: itave} In the averaged iteration process, each
  expectation value can be written as $\overline w_n(\rho^\Gamma,
  B_1)= \trace(A_n B_1 B_1^\dag)$, for the starting operator
  $B_1$. The sequence of operators $\{A_n\}_n$, that only depend on
  the given input operator $\rho^\Gamma$ with $\trace(\rho^\Gamma)=1$,
  is recursively defined by   
  \begin{eqnarray}
    \nonumber
    A_n&\equiv&A_n(\rho^\Gamma) \define A_{n-1} \\  \label{eq: recave} &+&
    \frac{1}{d} \left[ \trace(A_{n-1}) (\rho^\Gamma)^2\! -\! (\rho^\Gamma
    A_{n-1} + A_{n-1} \rho^\Gamma)\right]\!\!,  
  \end{eqnarray}
  starting with $A_1=\rho^\Gamma$, and $d$ denotes the dimension of
  the underlying Hilbert space.
\end{proposition}

\begin{proof}
The proposition follows by direct calculation of the averaged
expectation values. In order to perform this task, we employ the
identity that 
\begin{equation}
  \int dU\; d\trace(A U) \trace( B U^\dag)=\trace(AB),
\end{equation} 
holds for arbitrary operators $A, B$. This can be proven for example by
the Peter-Weyl theorem, see \textit{e.g.} Ref. \cite{folland}. (See
also note \footnote{The Peter-Weyl theorem states the identity $ \int
  dU \;d \braket{e_i| U | e_j} \braket{e_k| U^\dag|e_l}=\delta_{il}
  \delta_{jk}$, where $\{ \ket{e_i}_i\}$ is an arbitrary set of basis
  elements.}.) 
To actually compute the averaged expectation value $\overline w_n$,
given by Eq. \ref{eq: expave}, one uses the following \emph{strategy}:  

One starts with the average over the last chosen unitary $U_{n-1}$. Using the
recursion formula for the operator $B_n$, and the given identity, it
is possible to express the next integrand as the expectation value of
a witness in the step before with a new operator
$A_2=A_2(\rho^\Gamma)$, hence $\int d U_{n-1} \trace(\rho^\Gamma B_n
B_n^\dag) = \trace( A_2 B_{n-1} B_{n-1}^\dag)$. One continues with the
integration over the next unitary $U_{n-2}$, and exploits the same
trick again to obtain a new operator $A_3$. In this step, one needs to
be careful since the operator $A_2$ might not be normalized any more
$\trace(A_2)\not =1$. In the end, one uses this idea exactly $(n-1)$
times until one has performed all the integration and ends up with the final
operator $A_n$. For this operator one needs to compute the expectation
value with the first witness $B_1 B_1^\dag$, and obtains the final result
$\overline w_n =\trace(A_n B_1 B_1^\dag)$. 

Now that the strategy is fixed, one is left to perform the integration
over only one unitary in order to obtain the recursion
formula. Suppose the integrand is $\trace(A B_n B_n^\dag)$, where $A$
denotes an arbitrary hermitian operator. Using the recursion formula
for the operator $B_n$, given by Eq.~\ref{eq: recB}, the integrand is
expanded into   
\begin{eqnarray}
  \nonumber
  \trace(A B_n B_n^\dag)&=& \trace(A B_{n-1} B_{n-1}^\dag) \\
  \nonumber &&+ \trace(A)
  |\trace(\rho^\Gamma B_{n-1} U_{n-1})|^2 \\   
  \nonumber &&- \trace(A B_{n-1}
  U_{n-1}) \trace(\rho^\Gamma U_{n-1}^\dag B^\dag_{n-1})\\  
  & &- \text{c.c.} 
\end{eqnarray}
Since the operator $B_{n-1}$ is independent of the last unitary $U_{n-1}$,
one can directly perform the average over this last unitary. 
Each term can be integrated separately, and by using the given identity, this
results in 
\begin{eqnarray}
 &&\int d U_{n-1}\; \trace(A B_n B_n^\dag) \\   \nonumber &=&\!\!
 \trace\!\left( \!
   \left\{ A +\frac{1}{d}\left[ \trace(A) (\rho^\Gamma)^2
         \!-\!(\rho^\Gamma A + A \rho^\Gamma)\right] \!  \right\} \!
    B_{n-1} B_{n-1}^\dag \!\!\right)\!, 
\end{eqnarray}
so it can be expressed as $\trace(A^\prime B_{n-1}B_{n-1}^\dag)$,
where $A^\prime$ is given by the expression in the curly brackets,
that precisely gives the stated recursion formula of Eq. \ref{eq:
  recave}. This proves the proposition. 
\end{proof}

Next, one turns to the detection theorem itself. This theorem states
that any NPT entangled state will be detected in some step of the
averages iteration.  


\begin{theorem}[Detection for the averaged iteration] For any detectable
  operator $\rho^\Gamma \not \geq 0$ with $\trace(\rho^\Gamma)=1$, and
  any strictly positive operator $B_1 >0$, the averaged iteration
  process will always detect the state. 
\end{theorem}

\begin{proof}
In order to prove detection of an arbitrary state $\rho^\Gamma \not
\geq 0$, it is sufficient to show that the operators $A_n$, which
determine the average expectation value via Prop. \ref{prop: itave},
become negative definite at some point in the sequence. Since all
operators $A_n$ will necessarily commute with $A_1=\rho^\Gamma$, one
can easily identify the corresponding eigenvectors and just needs to
examine the behavior of the corresponding eigenvalues under the
iteration. Assume the following spectral decompositions of the operators
\begin{eqnarray}
  \rho^\Gamma &=& \sum_{i \in \mathcal{I}_+} \lambda^{+}_i \ket{v_i}\bra{v_i}
  + \sum_{j \in \mathcal{I}_-} \lambda^{-}_j \ket{v_j}\bra{v_j}, \\
  A_n &=& \sum_{i \in \mathcal{I}_+} a^{(n)}_i \ket{v_i}\bra{v_i}
  + \sum_{j \in \mathcal{I}_-} b^{(n)}_j \ket{v_j}\bra{v_j},
\end{eqnarray}
in which the index set $\mathcal{I}_{-}$ labels all the negative
eigenvalues of $\rho^\Gamma$, and $\mathcal{I}_+$ the strictly
positive eigenvalues. 

In order to show that all eigenvalues become negative at some point in
the iteration, we proceed as follows: First we show that all
eigenvalues on the negative subspace are decreasing exponentially with
the number of iterations $n$, whereas eigenvalues on the positive
semidefinite subspace can only increase linearly with $n$. This
already guarantees exponential fast divergence of
$\trace(A_n)$. Because of that, all eigenvalues on the positive
subspace will decrease exponentially fast as well, since the factor
$\trace(A_n)$ enters in the iteration formula, cf. Eq. \ref{eq:
  recave}. Combined with the fact that $B_1 > 0 $ this proves the
claim. In detail, we show by induction the following bounds: 
\begin{eqnarray}
  a_i^{(n)} &\leq& |\lambda_i^+| +
  (n-1)\frac{|\lambda_i^+|^2}{d}, \\
  b_j^{(n)} &\leq& -|\lambda_j^-|\left(
    1+\frac{|\lambda_j^-|}{d}\right)^{n-1}, 
\end{eqnarray}
for all $n\geq 1$. The induction start is trivial and we care about
the induction step $n \mapsto n+1$ only. Let us consider the positive
subspace first. In total we obtain the following sequence of inequalities 
\begin{eqnarray}
  \nonumber
  a_i^{(n+1)} &=& a_i^{(n)}\left( 1-\frac{2|\lambda_i^+|}{d} \right) +
  \frac{\trace(A_n)|\lambda_j^+|^2}{d} \\   
  \nonumber
  &\leq&  a_i^{(n)}\left( 1-\frac{2|\lambda_i^+|}{d} \right) +
  \frac{|\lambda_j^+|^2}{d} \\  
  \nonumber  
  &\leq& \left[ |\lambda_i^+| + (n-1) \frac{|\lambda_i^+|^2}{d}
  \right]\left( 1-\frac{2|\lambda_i^+|}{d} \right) +
  \frac{|\lambda_j^+|^2}{d} \\ 
  &\leq& |\lambda_i^+| + [(n+1)-1]\frac{|\lambda_i^+|^2}{d}.   
\end{eqnarray}
In the first line one has used the recursion formula of the operators
$A_n$ and the first inequality stems from the condition $\trace(A_n)
\leq \trace(A_1)=1$, that comes from the generic properties of the
iteration process, Prop.~\ref{prop: genprop} applied to the average
iteration process with $B_1=\mathbbm{1}$. In the next step, the
induction hypothesis is employed and one obtains the final result if
one upper bounds the term in the parenthesis. For the negative
subspace, one similarly obtains 
\begin{eqnarray}
  \nonumber  
  b_j^{(n+1)} &=& b_j^{(n)} \left( 1+\frac{2|\lambda_j^-|}{d}\right)
  + \frac{\trace(A_n)|\lambda_j^-|^2}{d} \\   
  \nonumber 
  &\leq& b_j^{(n)} \left( 1+\frac{2|\lambda_j^-|}{d}\right) +
  \frac{|\lambda_j^-|^2}{d} \\ 
  \nonumber   
  &\leq& b_j^{(n)} \left( 1+\frac{2|\lambda_j^-|}{d}\right) -
  b_j^{(n)} \frac{|\lambda_j^-|}{d} \\
  &\leq& -|\lambda_j^-|\left( 1+\frac{|\lambda_j^-|}{d}\right)^{(n+1)-1}.
\end{eqnarray}
Again, the recursion formula and $\trace(A_n)\leq 1$ were employed
first. The second inequality originates from the induction hypothesis,
since it allows to infer $|\lambda_j^-| \leq -b_j^{(n)}$. The
induction step finishes with another application of the induction
hypothesis for the last inequality. This concludes the proof of the theorem. 
\end{proof}

\section{Examples}
\label{sec: examples}

\begin{figure}[ht!!]
    \centering
    \includegraphics[scale=0.376,angle=-90]{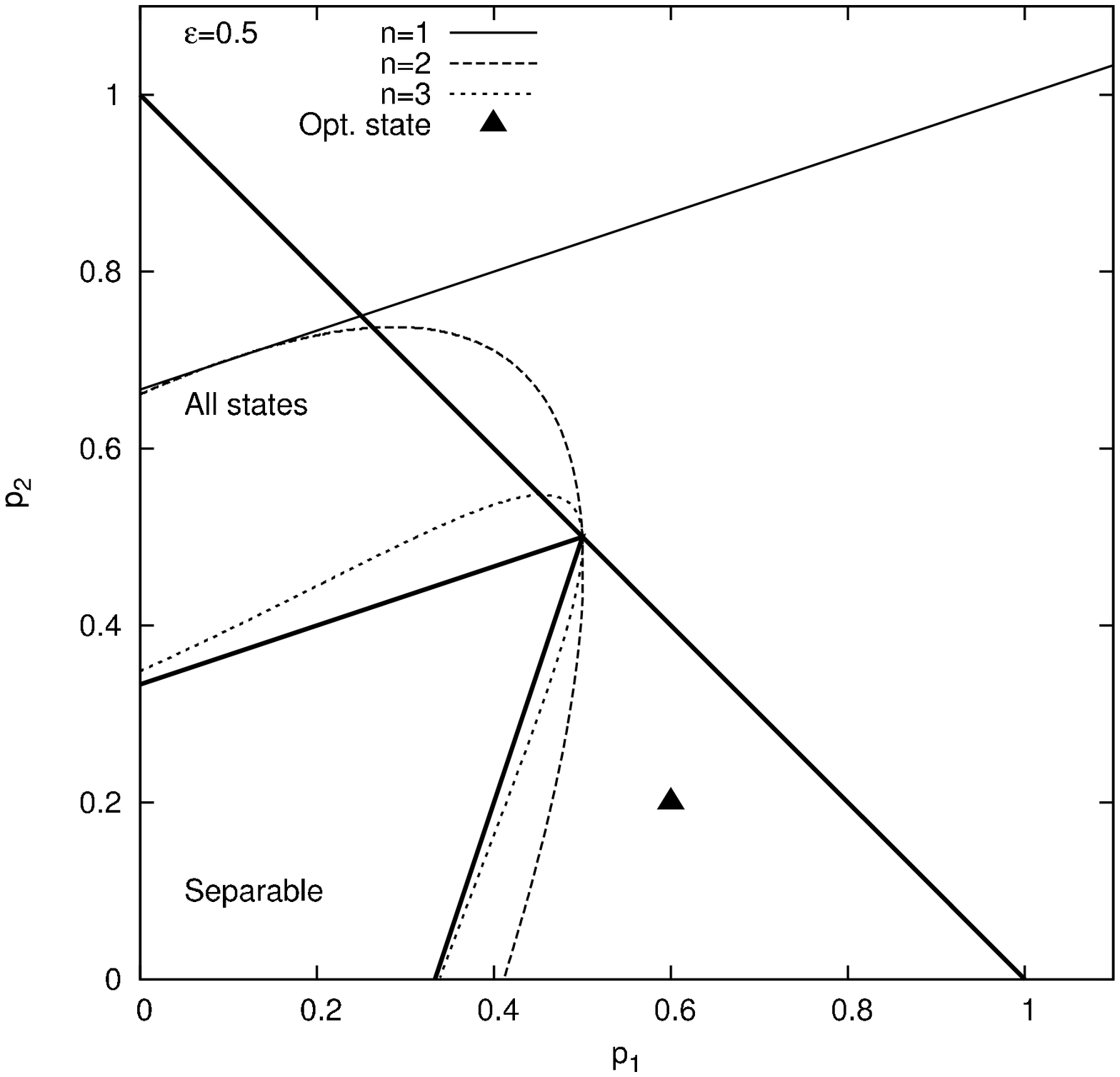} 
    \includegraphics[scale=0.376,angle=-90]{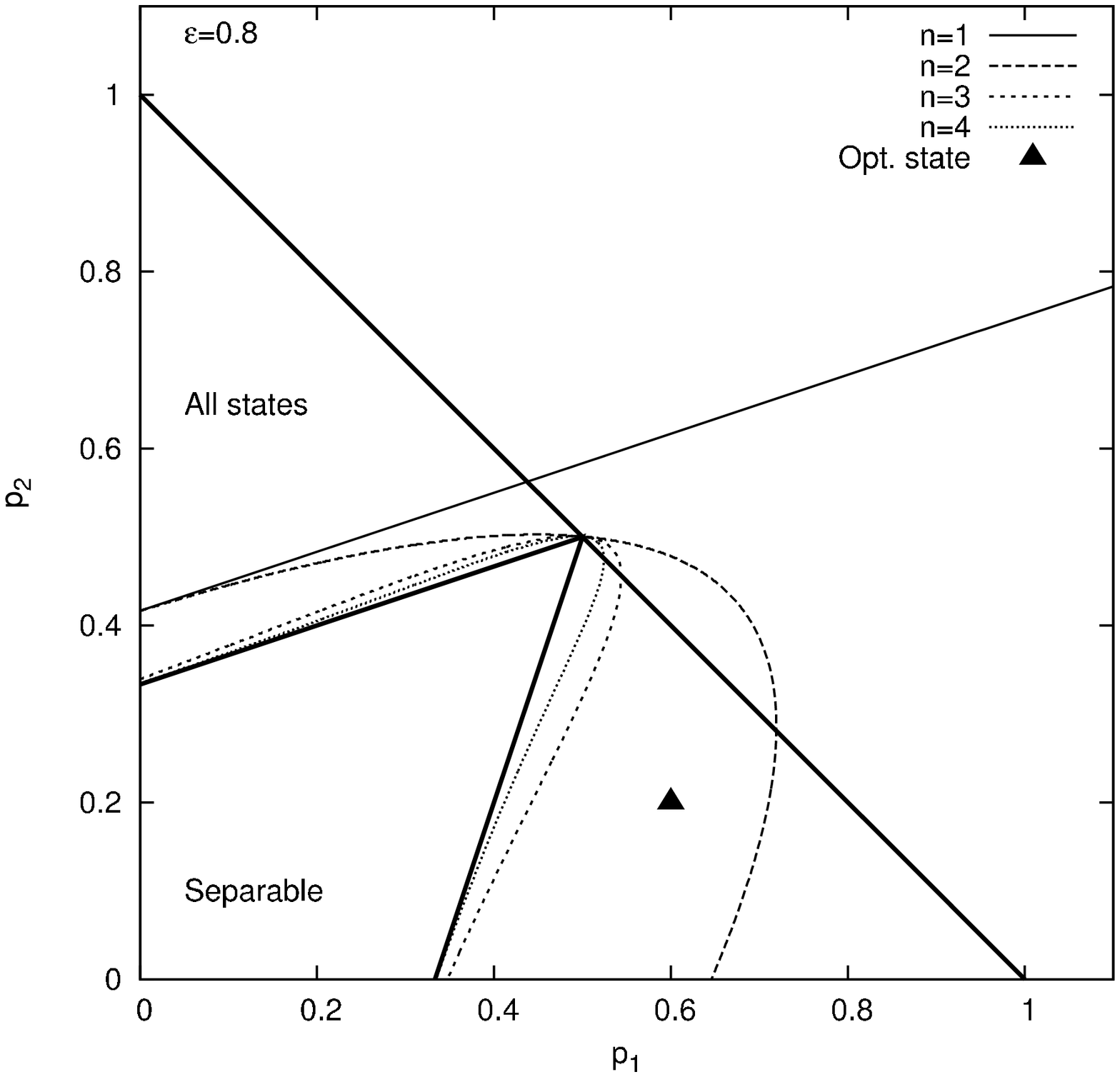}  
    \includegraphics[scale=0.376,angle=-90]{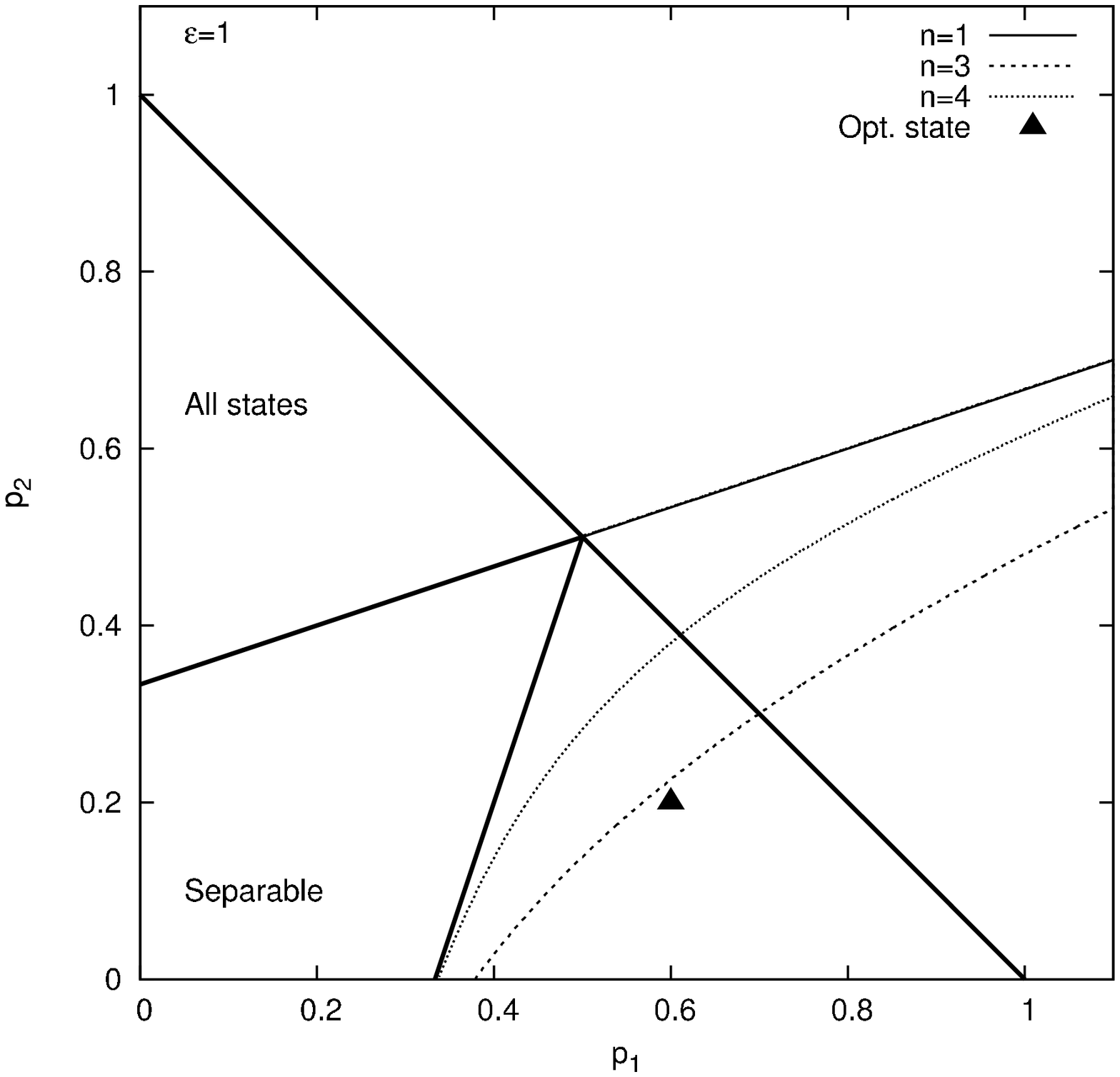}
    \caption{Optimized iteration: The set of physical states is given
      by the triangular shaped region connecting the extreme points
      $(0,1)$ and $(1,0)$ with the origin, while all separable states
      are described by the region inside. The black
      triangular, with coordinates $(0.6,0.2)$, symbolizes the target
      state. For the last case, $\epsilon=1$, each nonlinear
      entanglement follows the first linear entanglement witness (case
      $n=1$) and ``comes back'' on the other side.} 
    \label{fig: opt}
\end{figure} 

In order to visualize the effect of the iteration process---and to
provide a qualitative picture of the ``curvature'' of the corresponding
nonlinear entanglement witnesses---a simple two qubit example is
sufficient.

\subsection{Optimized Iteration}

To investigate the optimized iteration, we consider a starting witness 
that is a convex combination between an optimal entanglement witness 
and the identity operator; the parameter $\varepsilon$ describes the
corresponding mixedness of these two operators. This class of
entanglement witnesses with the corresponding operators $B_1$ is given by 
\begin{align}
  \label{eq: startingwitness}
  \WW &(\varepsilon)=\left(B_1 B_1^\dag\right)^\Gamma= \varepsilon
  \left(\ket{\psi^-}\bra{\psi^-}\right)^\Gamma +
  (1-\varepsilon)\frac{\mathbbm1}{4}, 
  \\ 
  B_1 &=\!\frac{1}{2}\left[\sqrt{1+3 \varepsilon} \ket{\psi^-}\bra{\psi^-} +
  \sqrt{1-\varepsilon} \left( \mathbbm{1}- \ket{\psi^-}\bra{\psi^-}
  \right)\!\right],    
\nonumber 
\end{align}
with $\ket{\psi^-}=\left(\ket{01}-\ket{10}\right)/\sqrt{2}$. In the
following we consider three different cases, the optimal entanglement
witness with $\varepsilon = 1$, and two slightly weaker witnesses with
$\varepsilon = 0.8$ and $\varepsilon = 0.5$ for different kinds of two
qubit states.  

First we investigate the nonlinear improvements of these witnesses for
a particular family of Bell-diagonal states given by 
\begin{equation}
  \rho=p_1 \ket{\phi^+}\bra{\phi^+} + p_2
  \ket{\phi^-}\bra{\phi^-} + 
  \left(1-p_1-p_2\right) \frac{\mathbbm{1}}{4},
\end{equation}
with the abbreviation
$\ket{\phi^{\pm}}=\left(\ket{00}\pm\ket{11}\right)/\sqrt{2}$ given in
a standard product basis, and $\mathbbm{1}/4$ denotes the totally
mixed state. Of course, only certain parameter pairs actually correspond to
physical states, since the corresponding operator must be positive
semidefinite in order to form a valid density operator. In the
Fig.~\ref{fig: opt}, the set of physical states corresponds to the
triangular shaped region. The convex set of separable states is
determined by the partial transpose and is given by the subset within
the set of physical states.  

The examples for the optimized iteration process are shown in
Fig.~\ref{fig: opt}, in which the target state is marked by the black
triangular. Note that this target state is on the ``wrong side'' of  
the state region, \ie, it is far from being detected by one of the
linear witness given by Eq.~\ref{eq: startingwitness}. As one can see,
the given target state is already detected after a few iterations. The
improvement, which characterizes the extend to which the entanglement
witness can be ``bent'' over to the other side of the state region,
depends on the original strength of the starting witness.  

Starting with a rather weak witness, the first example with
$\varepsilon=0.5$, already allows to curve the witness in such a way that
nearly all states on the other side are detected, while for the
slightly stronger witness, the second case with $\varepsilon=0.8$, the
first improvement is not yet enough to verify entanglement for the target
state. For the optimal entanglement witness, the last case
with $\varepsilon=1$, this task becomes the hardest, however even in this
case the target state is detected already after the third
iteration. For completeness, notice that starting with the identity
operator, $\varepsilon=0$, enables detection of any given target state
already after the first iteration. In addition note that if one
follows the optimized iteration process, the chosen target states 
is detected, but in general it will not be the case that one ends
up with an entanglement witness which is capable to witness all possible
states at once. 

\begin{figure}[t!!]
    \centering
    \includegraphics[scale=0.68,angle=0]{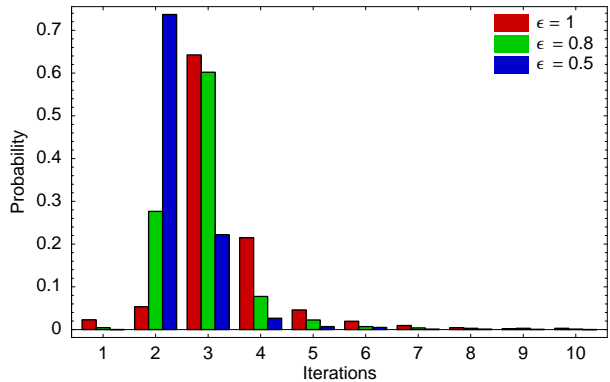}  
\caption{Optimized iteration: The probability of detecting a randomly 
chosen entangled state in the $n$-th step is shown, for the three values 
of $\varepsilon.$ See text for further details.}
\label{figoptstat}
\end{figure}

In a second example we investigate the number of required iterations
to detect a given target state. To this aim, we generated randomly (in
Hilbert Schmidt norm) entangled two-qubit states, and computed how
many iterations  are necessary, until they are detected starting from
each of the three witnesses given by Eq. \ref{eq:
  startingwitness}. The results are shown in
Fig.~\ref{figoptstat}. One can clearly see that most of the states are
detected after less than five iterations.   

\begin{figure}[t!!]
    \centering
    \includegraphics[scale=0.60,angle=0]{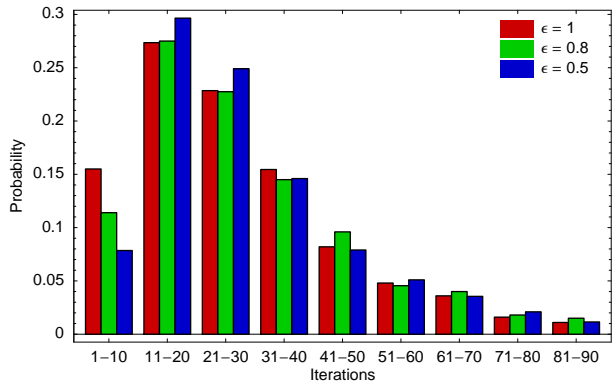}  
\caption{Random iteration: The probability of detecting a 
randomly chosen entangled state in certain intervals of iteration steps 
is shown, for the three values 
of $\varepsilon.$ See text for further details.}
\label{figrandomstat}
\end{figure}

\subsection{Random Iteration}
\label{sec: examples_random}

Let us now investigate the random iteration. For this aim, we consider
the three witnesses given by Eq.~\ref{eq: startingwitness} and
consider again randomly generated entangled states, and the number of
iterations which are required for an iteration. The results are
plotted in Fig.~\ref{figrandomstat}. On can see that most 
states are detected after hundred  iterations by randomly chosen
unitaries. In fact, we were unable to find an example of a state 
(also for other witnesses), which is not detected after five 
hundred iterations. This suggests that maybe also the random iteration
is complete, in the sense that each entangled state is detected after a 
finite number of steps.

\begin{figure}[ht!!]
    \centering
    \includegraphics[scale=0.376,angle=-90]{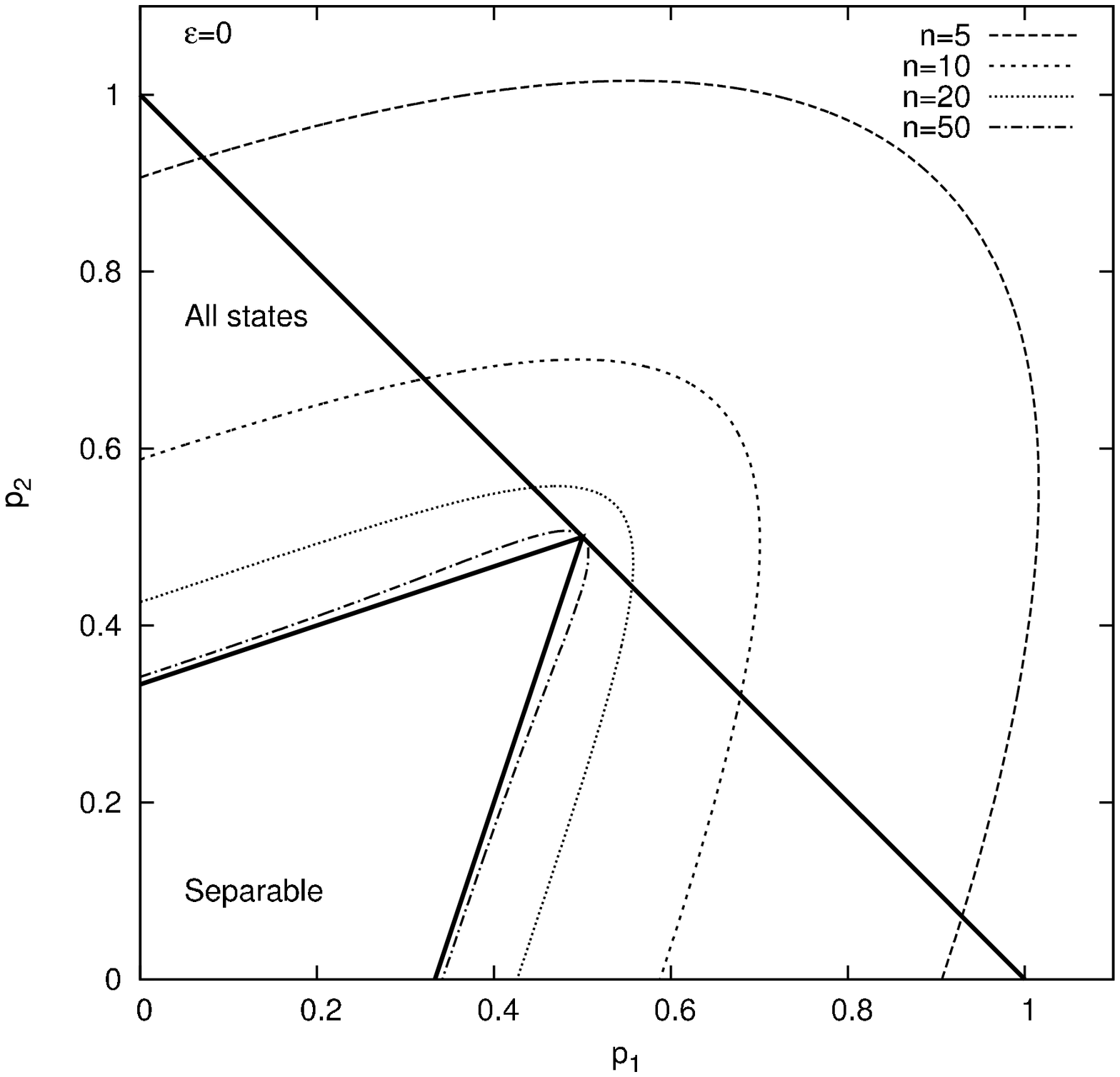} 
    \includegraphics[scale=0.376,angle=-90]{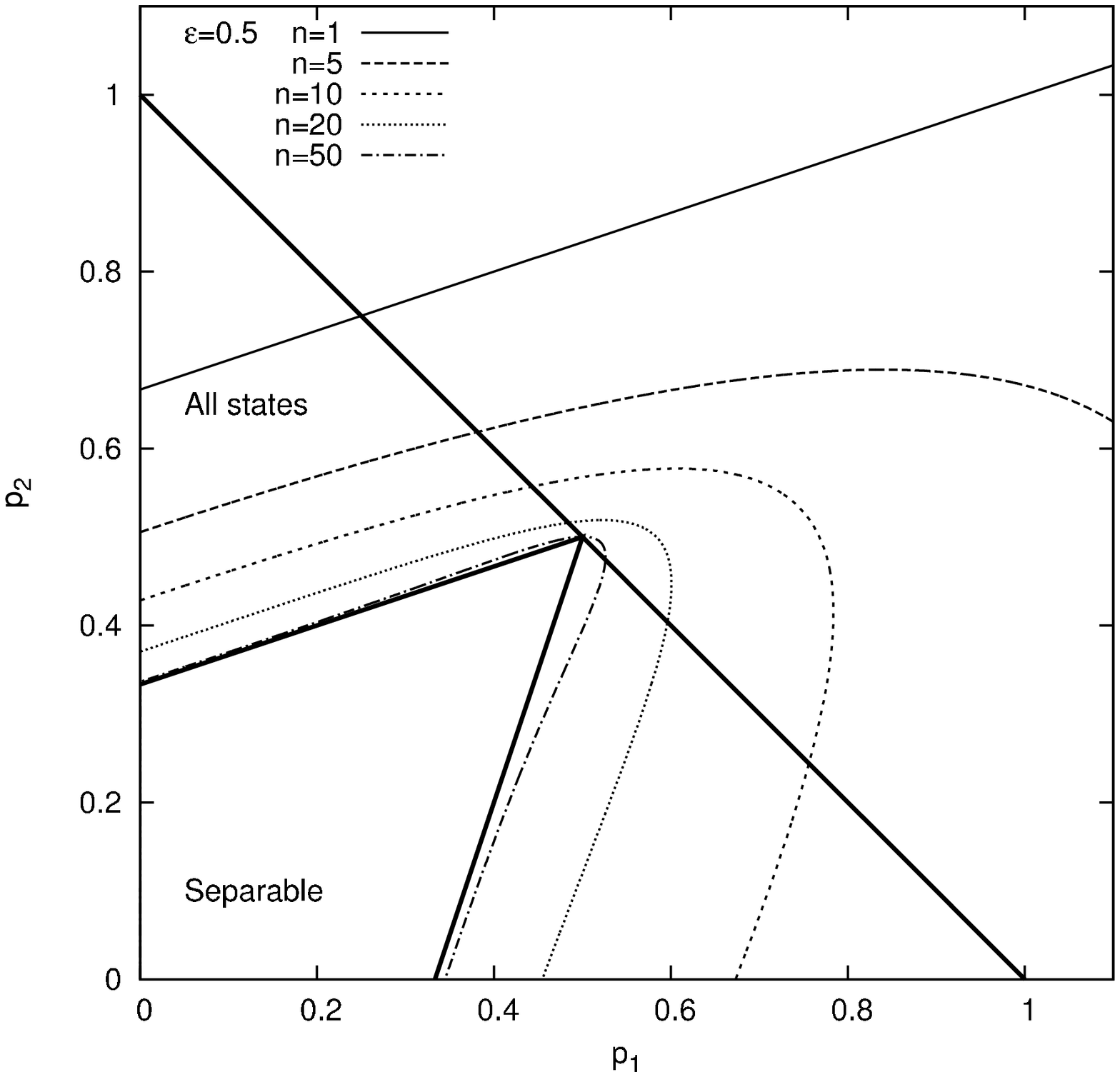}  
    \includegraphics[scale=0.376,angle=-90]{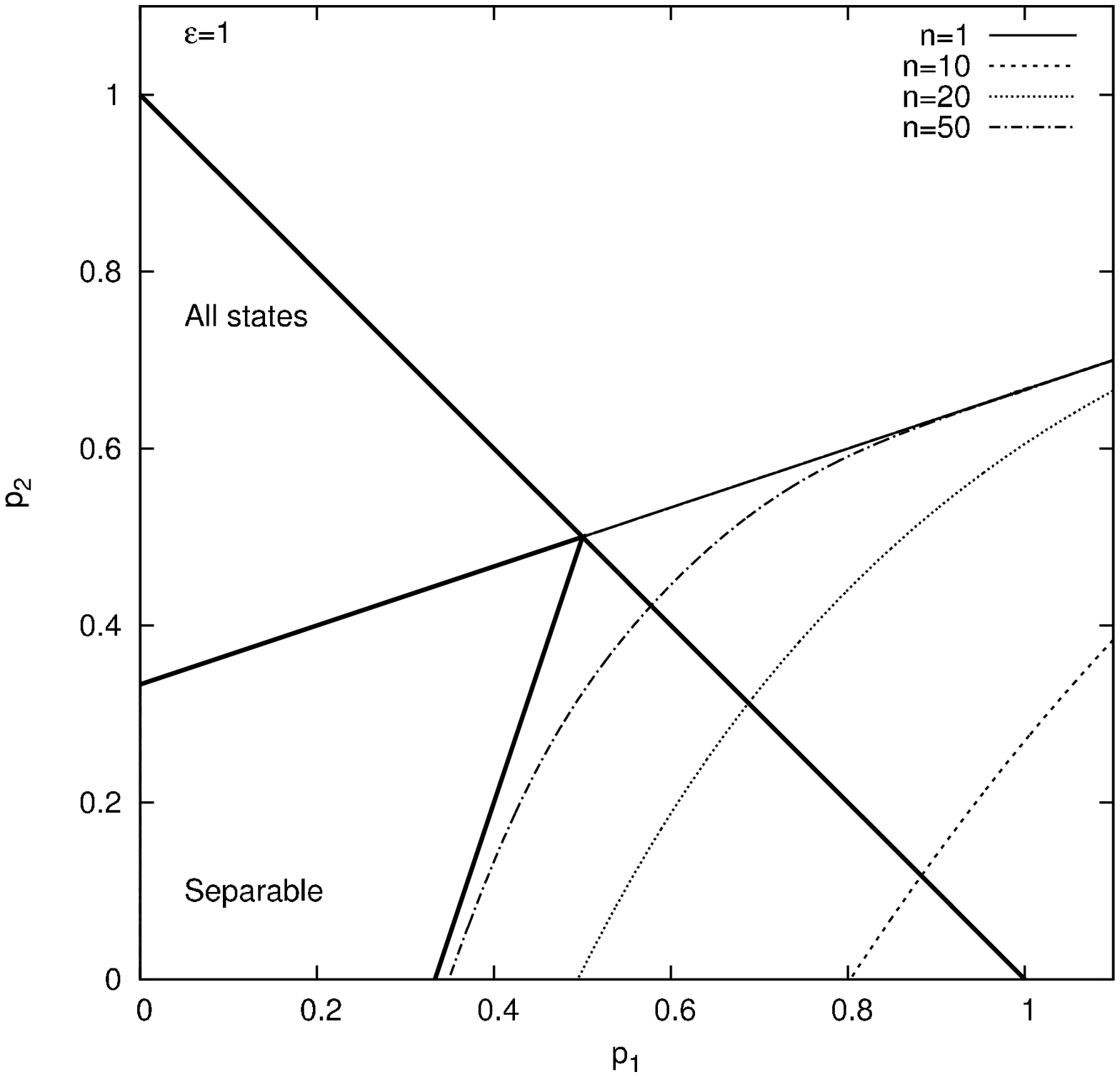}
    \caption{Averaged iteration: The set of physical states and the
      set of separable states are given as in optimized iteration,
      cf. Fig. \ref{fig: opt}. In the last case, $\varepsilon=1$, the
      detected states are again the ones which are already detected by
      the first, linear witness (case $n=1$), plus the states in the
      right region if the witness curves back.}
    \label{fig: ave}
\end{figure}

\subsection{Averaged Iteration}

Finally, Fig.~\ref{fig: ave} shows similar examples for the averaged iteration
process, in which the same starting witnesses are employed as for the
optimized sequence. As one can see, the region of detected states
increases with the number of considered iterations, however much more
iterations are needed in comparison to optimized iteration
process. For the first example, $\varepsilon=0$, the identity operator
is used as the starting operator. Using the averaged iteration method,
the part of detected states is symmetric with respect to the upper and
lower state region, which shows that the corresponding witness is
improved equally in all possible directions. This symmetry breaks of
course if one uses one of the asymmetric entanglement witnesses to
start with.

\section{Extensions}
\label{sec: conclusion}

Although the iteration process and its corresponding results were
solely discussed for the PPT criterion, the method directly
generalizes to other trace-preserving, positive but not completely
positive maps $\Lambda$; for example to the reduction criterion and
various extensions of it \cite{reduction, piani}, or the Choi-map
\cite{choi}. All entangled states which violate the corresponding
condition $\rho^\Lambda \equiv \text{id} \otimes \Lambda(\rho) \geq 0$
can be detected by the iteration method. In order to obtain the
entanglement witness in the usual sense, one employs the adjoint map
$\Lambda^\dag$, defined by the property that   
\begin{equation}
\trace( \Lambda(X) \; Y)=\trace(X\; \Lambda^\dag(Y)), \;\; \forall X,Y.
\end{equation}
Therefore the entanglement witness becomes $\WW \define \text{id}\otimes
\Lambda^\dag (BB^\dag)$, which represents the general connection
between the entanglement witness and the iteration operator $B$.

Finally, due to the isomorphisms studied by de Pillis,
Jamio{\l}kowski, and Choi \cite{pillis-67, jamiolkowski, choi-82},  we
can find for a given witness always a positive, but not completely
positive map $\Lambda$, and a state $\ket{\psi}$ such that the witness
can be written as 
\begin{equation}
  \WW = \text{id}\otimes \Lambda^\dagger (\ketbra{\psi}{\psi}),
\end{equation}
see Ref.~\cite{nlwit1,nlwit2} for details. This is the same structure
as the witness in Eq. \ref{basicwitness}, hence all the results can be applied.

\section{Conclusion}

In conclusion, we provided a sequence of  nonlinear entanglement witnesses, 
defined as functionals on the set of quantum states, which necessarily are 
non-negative for all separable states. 

Two particular iterations were investigated in more detail. In the
optimized iteration one tries to optimize the improvement, given by
the nonlinear term, according to a given preselected target
state. Form the geometric picture of entanglement witness,
cf. Fig.~\ref{fig: motivation}, this process corresponds to the task
that one likes to ``curve'' a given witness along a certain
direction. By contrast, the averaged iteration deals with the exact
opposite case; one tries to improve the witness along all possible
directions. The main result of the manuscript is that both iteration
methods are successful: Any entangled state, detectable by the
corresponding positive but not completely positive map, is also
detected by the corresponding sequence of nonlinear entanglement
witness, as long as the starting witness does not act onto a
completely different subspace.


There are several open questions which deserve further study. First,
it remains open whether a similar detection statement holds for the
random iteration process, as suggested by the numerical example of
Sec.~\ref{sec: examples_random}. Second, it might be desirable to
extend the iteration idea beyond a certain positive but not completely
positive map; however this seems to require some extra knowledge about
the structure of the quantum states applied to some particular map. In
addition, it is tempting to ask whether similar ideas translate to
other known entanglement criteria, which do not rely on a positive but
not completely positive maps, but which can be considered in an
entanglement witness form, \textit{e.g.} the computable cross norm or
realignment criterion \cite{crossnorm,realignment}. Finally, it
remains to clarify the possible connection between the averaged
iteration process and the spectrum estimation idea from Ref.~\cite{keyl}.

\section{Acknowledgments}

We would like to thank Marco Piani, Hauke H\"aseler and Xiongfeng Ma for 
interesting discussions. This work was funded by the European Union 
(OLAQUI, QAP, QICS, SCALA, SECOQC),
the NSERC Innovation Platform Quantum Works and the NSERC Discovery grant
and the FWF (START prize).

\bibliographystyle{apsrev}


\end{document}